\documentclass[draft,
]{svjour2}
\usepackage{amsmath,amssymb,latexsym} 
\def\be{\begin{equation}}
\def\ee{\end{equation}}
\def\bea{\begin{eqnarray}}
\def\eea{\end{eqnarray}}
\def\nn{\nonumber}

\begin{document}

\title{An Elementary Derivation of the Five Dimensional Myers-Perry Metric}

\authorrunning{J. -J. Peng }

\author{Jun-Jin Peng }

\institute{College of Physical Science and Technology, Central China Normal
University, Wuhan, Hubei 430079, People's Republic of China \\
\email{pengjjph@163.com}}
\date{}

\maketitle

\abstract{We derive the five dimensional Myers-Perry metric via an elementary
method to solve the vacuum Einstein field equations directly. This method
firstly proposed by Clotz  is very simple since it merely involves four components
of Ricci tensor and only requires us to deal with some equations without second
derivatives' terms when the metric ansatz is assumed to take an appropriate form.}

\section{Introduction}

In the last few years an increasing effort has been devoted to the study of black
holes in higher dimensional gravity \cite{RCMP} \cite{REHR} \cite{EEMR} \cite{REMRLL}
\cite{HARM} \cite{HARMPO} \cite{Aliev} \cite{SQW} for various reasons. Among them,
the most direct and important ones stem from the motivation by string and membrane
theories. In the context of the string theory, which is thought to be a major
candidate for the quantum theory of gravity and the unification of all interactions,
black holes in higher dimensions are consistent with its mathematical requirement of
higher dimensions. They may not only play a key role in the study of dynamics in higher
dimensions but also serve to advance our understanding of the compactification mechanisms.
In particular, microscopic black holes may be very useful to test some novel predictions
of the string theory. On the other hand, in the brane world scenario, if the assumption
that there exist extra dimensions in the universe holds true, it is possible to produce
higher dimensional mini black holes in future high energy colliders  or observe them in
the universe \cite{BRANE} \cite{BRANEGT} \cite{BRANEKA}.

In light of the above, it becomes clear that the study of finding exact solutions in higher
dimensional gravity is of great importance. However, it is notorious that this task is very
difficult, especially to discover the ones in higher dimensional general relativity, compared
with the case in four dimensions, where several systematic techniques to generate stationary
and axisymmetric solutions have been developed by some authors \cite{ERNS} \cite{ERNSB}
\cite{STEPH} and a lot of interesting solutions are presented, such as the unique rotating
Kerr black hole for a given mass and angular momentum \cite{KERR} and the charged Kerr-Newman
black hole \cite{KERRN}. Till now, there are still few known exact higher dimensional solutions
in general relativity. The first exact rotating one of vacuum Einstein field equations was
given by Myers and Perry (MP) in 1986 \cite{RCMP}. This solution is the higher dimensional
generalization of the classical Kerr black hole and has a spherical horizon topology.
Nevertheless, in the case of the five dimensional pure gravity, MP black hole is not unique.
For a given mass and angular momentum, Emparan and Reall discovered a black ring
solution in vacuum, whose horizon topology is $S^1 \times S^2$ \cite{REHR}. Then
another general regular black ring solution with two angular momenta was also found in
\cite{AAPS}. Quite recently, in \cite{LUMP}, the authors presented a new five dimensional
black hole solution with the horizon of distorted lens spaces.

As a result, systematic generation-techniques of higher dimensional solutions have to be
developed. Recently, a new technique was applied to study the stationary and axisymmetric
solutions in the arbitrary dimensional pure gravity \cite{HARM}. this technique generalizes
the Papapetrou form of the metric for stationary and axisymmetric solutions in four dimensions,
and furthermore extends the work on Weyl solutions in four and higher dimensions, which gives
the five dimensional MP solution. In fact, several other generalized methods were also used
to reproduce the five dimensional MP metric. In \cite{AAPO}, a four dimensional technique
named the inverse scattering method was extended to derive this metric. Besides, applying a
sequence of $SO(2,1)$ transformations on the five dimensional Schwarzschild metric, the authors
obtain the five dimensional MP solution, too \cite{SGIO}.

In this short paper, we will employ an elementary method that only requires us to solve the
vacuum Einstein field equations directly to derive the five dimensional MP
metric. This method was first used to reproduce Kerr solution by Klotz \cite{KLOTZ}. Then
it was generalized to derive the Kerr-Newman black hole \cite{GUIY}. According to this
method, when the metric ansatz is assumed to have an appropriate form, we only need solve
two equations which merely refer to four components of Ricci tensor to derive the five
dimensional MP solution. Our paper goes as follows. Section two is the main part, where we
shall explicitly deduce the five dimensional MP black hole solution. In section three, a
simple conclusion is presented.

\section{calculations}
\label{bhgr}

Bearing in mind that our prime aim is to derive the five dimensional MP black hole which is
the higher dimensional generalization of the Kerr solution in four dimensions, the metric
ansatz extended from the one in \cite{KLOTZ} is assumed to take the general form
\bea
ds^2 &=& [h(x,\theta)-1]d\tau^2
+ \Sigma(x,\theta)\big(dx^2+d\theta^2+p(\theta)d\phi^2+q(\theta)\psi^2 \big) \nn \\
&&+ 2ap(\theta)d\tau d\phi
+ 2bq(\theta)d\tau d\psi - \big(b p(\theta)d\phi+a q(\theta)d\psi \big)^2 \label{GMetr} \, ,
\eea
where parameters $a$ and $b$, which can be absorbed by the functions $p(\theta)$ and $q(\theta)$
respectively, are two constants inserted for dimensional reasons. Compared with the ansatz
in \cite{KLOTZ}, the last term in Eq. (\ref{GMetr}) arises from the two angular momenta.
We perform a coordinate transformation
\be
d\tau = dt + apd\phi + bqd\psi \label{coortra} \, ,
\ee
and also assume
\be
\Sigma = F(x) - a^2p-b^2q \label{ftra} \, .
\ee
Here the assumption that $F(x)$ is only a function of the coordinate $x$ may be too strong.
It would be avoided. However, owing to this restriction on $F(x)$, the field equations are
easy to separate, which simplifies the calculation by a long way. Without this assumption,
it is nearly impossible for the calculations to proceed.

Now, let's calculate the Ricci tensor of metric (\ref{GMetr}) in the coordinate system
$(t,x,\theta,\phi,\psi)$. Its nine non-zero components are $R_{tt}$, $R_{t\phi}$,
$R_{t\psi}$, $R_{xx}$, $R_{x\theta}$, $R_{\theta\theta}$, $R_{\phi\phi}$, $R_{\phi\psi}$
and $R_{\psi\psi}$. With aids of the two functions
\bea
Q(\theta) &=& p + q \, ,  \\
\Delta (x,\theta) &=& (F-a^2Q)(F-b^2Q)  \nn \\
           &&- h(F-a^2p-b^2q)[F-(a^2+b^2)Q] \, ,
\eea
where the function $\Delta$ is easy to construct once we note that a lot of components of
Christoffel brackets contain it in their denominators, the component $R_{x\theta}$ can be
expressed as
\be
R_{x\theta} = A(x,\theta)\Delta_{,x\theta} + B(x,\theta)\Delta_{,\theta}
+ C(x,\theta)Q_{,\theta} \, . \label{Rxtheta}
\ee
In Eq. (\ref{Rxtheta}), functions
$A$, $B$ and $C$ are too complicated to list. Accordingly, in order to keep $R_{x\theta} = 0$,
the simple and also final assumption is that $\Delta$ is only the function of the variable
$x$. With this assumption, we can get $C = 0$ or  $Q_{,\theta} = 0$. Taking the latter into
consideration, $Q$ and $h$ have the simplified forms
\bea
Q &=& const = k \, , \nn \\
h &=& \frac{-\Delta(x) + (F - ka^2)(F - kb^2)}{(F-a^2p-b^2q)[F-k(a^2+b^2)]}
 \, . \label{Qhassum}
\eea

Up to the present, there are still three functions to be determined. If we directly solve the
other non-zero components of the Ricci tensor, the second derivatives to $x$ and $\theta$
nearly make this plan impossible. For the sake of avoiding this obstacle, we take into account
the following equation
\be
abkR_{tt} - bR_{t\phi} - aR_{t\psi} =0 \, , \label{Rtphipsi} \\
\ee
whose merit is that all the terms including the second derivatives disappear. Introduce a new
independent variable $\sigma$, which satisfies
\be
d\sigma = \Delta^{\frac{1}{2}} dx \, , \\
\ee
and a new function $H(x)$ to rewrite $\Delta(x)$ as the form
\be
\Delta = (F-ka^2)(F-kb^2) - H[F - k(a^2+b^2)] \, . \\
\ee
Hence Eq. (\ref{Rtphipsi}) is simplified as
\be
F_{,\sigma}^2 - \frac{H_{,\sigma}F_{,\sigma}}{H}(F - kb^2) + \frac{p_{,\theta}^2}{p(p-k)}
-\frac{H_{,\sigma}F_{,\sigma}}{H}(b^2-a^2)p = 0 \, . \label{HFP}
\ee
Since $F$ and $H$ are only the functions
of $\sigma$ while $p$ relates to the variable $\theta$, we must have
\bea
p_{,\theta}^2 &=& p(\alpha p+\beta)(p-k)  \, ,  \label{psolu} \\
F_{,\sigma}^2 &=& \frac{\alpha}{b^2-a^2}F - \frac{\alpha kb^2}{b^2-a^2} - \beta \, , \qquad
\frac{H_{,\sigma}F_{,\sigma}}{H} = \frac{\alpha}{b^2-a^2} \, , \label{FHsolu}
\eea
in which $\alpha$ and $\beta$ are arbitrary constants. When $\alpha = 0$ and $\beta = -l^2$,
solving Eqs. (\ref{psolu}) and (\ref{FHsolu}), $p$, $q$, $\Sigma$ and $h$ take the general forms
\bea
p &=& k \sin^2\Big(\frac{l}{2}\theta \Big) \, , \qquad
q = k \cos^2\Big(\frac{l}{2}\theta \Big) \, , \\
\Sigma &=& l\sigma +n - k\Big[a^2\sin^2\Big(\frac{l}{2}\theta \Big)
+  b^2\cos^2\Big(\frac{l}{2}\theta \Big)  \Big] \, ,  \\
h &=& \frac{2M}{l\sigma + n-ka^2\sin^2(l\theta /2)-kb^2\cos^2(l\theta /2)} \, ,
\eea
respectively, where $l$, $n$ and $M$ are arbitrary constants. If we substitute them into Eq.
(\ref{GMetr}), it is an easy matter to check that Eq. (\ref{GMetr}) satisfies the vacuum
Einstein equations. In particular, adopting
\be
k = 1 \, ,  \qquad l = 2 \, , \qquad n = a^2+b^2 \, , \qquad \sigma = \frac{r^2}{2} \, ,
\ee
we obtain the five dimensional MP metric in the Boyer-Lindquist coordinates
\bea
ds^2 &=& \Big(\frac{2M}{\Sigma}-1\Big)d\tau^2
+ \Sigma \Big(\frac{r^2}{\Delta}dr^2 + d\theta^2 + \sin^2\theta d\phi^2
+\cos^2\theta d\psi^2 \Big)  \nn \\
&& +2a\sin^2\theta d\tau d\phi +2b\cos^2\theta d\tau d\psi
  -(b\sin^2\theta d\phi+ a\cos^2\theta d\psi)^2\, ,
\eea
where
\bea
d\tau &=& dt+a\sin^2\theta d\phi +b\cos^2\theta d\psi \, ,  \\
\Sigma &=& r^2 + a^2\cos^2\theta + b^2\sin^2\theta \, , \\
\Delta &=& (r^2 + a^2)(r^2 + b^2) -2Mr^2 \, .
\eea
On the other hand, when $\alpha \neq 0$, the solutions got from Eq. (\ref{HFP}) does not
satisfy the Einstein field equations in vacuum. This means that the solution of Eq. (\ref{GMetr})
is unique on the assumption that Eqs. (\ref{ftra}) and (\ref{Qhassum}) hold.

\section{Final remarks}

In this paper, we have reproduced the five dimensional MP metric by an elementary method. Since
this method only requires us to solve the equation $R_{x\theta} = 0$ and Eq. (\ref{Rtphipsi})
with respect to the four components of the Ricci tensor $R_{x\theta}$, $R_{tt}$, $R_{t\phi}$
and $R_{t\psi}$, it is simple although the calculation is a little complicated and we have presented
the assumptions that the metric ansatz takes the general form (\ref{GMetr}) and the
functions $\Delta$ and $F$ only depend on the variable $x$ to make the Einstein field equation
simplified. Moreover, this method might be general. In our future work, we will extend it to
generate other neutral or charged solutions in higher dimensions.

\section*{Acknowledgments}

I would like to thank Professor S. Q. Wu for enlighten advices and useful discussions.
This work was supported in part by the Natural Science Foundation of China under Grant
No. 10675051.

\end{document}